\def\lf{\hfil\break}
\def\bull{\lf$\bullet$ }
\def\eq#1{Eq.~(\ref{#1})}
\def\D{{\cal D}}
\def\pcr{\nonumber\\}
\def\half{{\scriptstyle {1 \over 2}}}

\newcommand{\beq}{\begin{equation}}
\newcommand{\be}{\begin{equation}}
\newcommand{\eeq}{\end{equation}}
\newcommand{\ee}{\end{equation}}

\newcommand{\bea}{\begin{eqnarray}}
\newcommand{\eea}{\end{eqnarray}}
\newcommand{\bean}{\begin{eqnarray*}}
\newcommand{\eean}{\end{eqnarray*}}

\documentclass[12pt]{revtex4}
\begin{document}

\title{Stationary phase in Coherent State Path Integrals}
\author{Chris Ray}
\affiliation{Department of Physics, Saint Mary's College, Moraga, CA 94575}
\email{cray@stmarys-ca.edu}
\author{Giulio Ruffini}
\affiliation{Starlab SL, Barcelona\\ Edifici de l'Observatori Fabra\\ Muntanya del Tibidabo Cami de l'Observatori s/n, 08035 Barcelona, Spain}
\email{giulio@starlab.net}

\date{\today}

\begin{abstract}
In applying the stationary phase approximation to coherent state path integrals a difficulty occurs; there are no classical paths that satisfy the boundary conditions of the path integral.  
Others \cite{K1},\cite{IJMP} have gotten around this problem by reevaluating the action.  
In this work it is shown that it is not necessary to reevaluate the action because the stationary phase approximation is applicable so long as the path, about which the expansion is performed, satisfies the associated Lagrange's equations of motion.  
It is not necessary for this path to satisfy the boundary conditions in order to apply the stationary phase approximation.

\end{abstract}

\maketitle
\vfill\eject

%\tableofcontents

%%%%%%%%%%%%%%%%%%%%%%%%%%%%%%%%%%%%%%%
\section{introduction}
The propagator from the coherent state $A$ to the coherent state $B$ over a time interval $\tau$ can be written \cite{K2} as a path integral over
coherent state paths from $A$ to $B$.
\begin{equation}\label{propPI}
\langle {B} \vert U(\tau) \vert {A} \rangle =\int_A^B \D L\  e^{iS[L]/\hbar}
\end{equation}
The integral is over all complex paths $L(t)$ such that $L(0)=A$ and $L(\tau)=B$, and the action for the path $L(t)$ is defined as
$$
S[L]\equiv \int_0^\tau dt \left[ i{{\hbar} \over 2}\left(\dot L L^* - L \dot L^*\right) - \langle {L} \vert {H} \vert {L} \rangle \right]
$$

If the expectation value of the hamiltonian ($H$) is truncated to second order in $L$ it is expected, from experience with path integrals \cite{FeynmanAndHibbs}, that by expanding about the ``classical path'' from $A$ to $B$, the path integral will be reduce to a gaussian path integral with trivial boundary conditions.
This procedure is the adaptation of the  stationary phase approximation to path integrals.
Unfortunately this cannot be done directly.
The difficulty is that the action integral is linear in $\dot L$, and thus the equation of motion for the classical path is first order.
Consequently a solution to the equation of motion will not have sufficient freedom to choose both end points.  
Thus, in general, there is no classical path that matches the boundary conditions.

Klauder suggests circumventing this problem by including a $\dot L^2$ term in the action that vanishes in the path integral limit \cite{K1}.
This leads to a second order equation of motion for the classical path and thus classical paths that can match the boundary conditions at both ends.

More recently Shibata and Takagi have suggested \cite{IJMP} that the failure of the stationary phase approximation for the coherent state path integral indicates that the action, quoted above, for the coherent state path integral is incorrect, and that it is necessary to start over with a discrete-time formalism in each case in order to get the correct result.

This work demonstrates that there is no need to reevaluate the coherent state action.
A coherent state path integral can be approximated by the stationary phase approximation without altering the action or starting over with a discrete-time formalism.

This paper will first find the propagator using the coherent state path integral, and show that this the propagator is correct.
Then the apparent conflict between this result and the work of Klauder and Shibata will be addressed.

%%%%%%%%%%%%%%%%%%%%%%%%%%%%%%%%%%%%%%%%%%

\section{The Extreme-Path Solution}

Consider then the general, truncated to second order, expectation value of the Hamiltonian, written as 
\begin{equation}\label{truncatedH}
\langle {L} \vert {H} \vert {L} \rangle \approx \hbar\left(\omega LL^* + fL+f^*L^*  \right)
\end{equation}
with $\omega $ real and $f$ a complex function of time.
With this truncation, the action can be written in the following form.
$$
S[L] = \int_0^\tau dt\ {\cal L}[L(t),L^*(t)] \equiv S[L,L^*]
$$
 with
$$
{\cal L}[L(t),M^*(t)] \equiv \hbar\left[{ i \over 2}\left(\dot L M^* - L \dot M^*\right) -\omega LM^* - fL - f^*M^*  \right]
$$

Considering $L$ and $M$ as independent functions, the associated Lagrange's equations of motion are
\begin{eqnarray}\label{LeEqMo}
i\dot L &=& \omega L + f^* \pcr
-i\dot M^* &=& \omega M^* + f
\end{eqnarray}
We know that deviations from paths satisfying \eq{LeEqMo} induce no first order deviation in the action since these paths are stationary points of the action.
For this reason, paths that satisfy \eq{LeEqMo} will be called extreme paths in this paper.
Indeed we find after an integration by parts that there are no linear terms in the integral over the deviations.
\begin{eqnarray}\label{boundaryTerm}
S[L_0+\delta L, M_0^*+\delta M^*]
&=& S[L_0, M_0^*] \pcr
&+& { i \over 2}\hbar\left[M_0^*\delta L-L_0\delta M^*\right]_0^t \cr
&+& \hbar\int_0^\tau dt \left[{ i \over 2}\left(\dot {\delta L} \delta M^* - \delta L \dot {\delta M}^*\right) -\omega \delta L\delta M^*\right] 
\end{eqnarray}
Where $M_0$ and $L_0$ satisfying the equations of motion \eq{LeEqMo}.  

Oddly, $L$ and $L^*$ can be expanded about different extreme paths and the resulting integral in the action will still have no linear terms.  
Note also that this lack of linear terms has nothing to do with the boundary conditions, and depends only on the fact that the paths satisfy the equations of motion.

While it is true that $L$ and $L^*$ can be expanded about different paths and still get the simplification of \eq{boundaryTerm}, in the path integral we need the action, $S[L,L^*]$, for a single path $L$.
Thus for present purposes it must be demanded that $(M_0^* + \delta M^*)^* = L_0 + \delta L$ and thus that
\be
M_0 + \delta M = L_0 + \delta L .
\ee
It is apparent then that $\delta L$ and $\delta M$ are not really independent.  Which is as it should be since this action is in a path integral over a single path.
For a given choice of $M_0$ and $L_0$ we can write $\delta M$ in terms of $\delta L$.
$$\delta M = L_0 - M_0 + \delta L$$ 
This can be used to write the action in terms of a single function, $\delta L$.
The following is found, after an integration by parts and using the fact that $L_0$ and $M_0$ are solutions to
\eq{LeEqMo}.
\begin{eqnarray}
S[L,L^*]
&=& S[L_0+\delta L, M_0^*+\delta M^*] \pcr
&=& S[L_0, M_0^*] \pcr
&&+ { i \over 2}\hbar[M_0^*\delta L-L_0\delta M^*]_0^t + { i \over 2}\hbar[L_0^*\delta L - M_0^*\delta L ]_0^t \pcr
&&+ \hbar\int_0^\tau dt \left[{ i \over 2}\left(\dot {\delta L} \delta L^* - \delta L \dot {\delta L}^*\right) -\omega \delta L\delta L^*\right] \pcr
&=& S[L_0, M_0^*] \pcr
&&+ { i \over 2}\hbar[ L_0^*  \delta L-L_0\delta M^* ]_0^t \pcr
&&+ \hbar\int_0^\tau dt \left[{ i \over 2}\left(\dot {\delta L} \delta L^* - \delta L \dot {\delta L}^*\right) -\omega \delta L\delta L^*\right] \nonumber
\end{eqnarray}

Now comes the nice part.  The first two terms do not depend on $\delta L$, just the boundary conditions and the extreme paths.
Thus these terms may be brought outside the path integral over $\delta L$ in the path integral representation of the propagator, \eq{propPI}.
The last term, on the other hand,  is the action associated with a harmonic oscillator hamiltonian $\hbar \omega a^\dagger a$, where $a$ is the annihilation operator ($a\vert L\rangle = L\vert L\rangle$).
Thus
\begin{eqnarray}\label{prop1}
\langle {B} \vert U(\tau) \vert {A} \rangle
&=& \exp\left\{ {i \over \hbar}S[L_0, M_0^*] \right\} \pcr
&\times& \exp\left\{- { 1 \over 2}[ L_0^*  \delta L-L_0\delta M^* ]_0^\tau \right\} \pcr
&\times& \int_{A-L_0(0)}^{B-L_0(\tau)} \D [\delta L]\  e^{iS_{\rm HO}[\delta L]/\hbar}
\end{eqnarray}

The third term is the propagator for a harmonic oscillator hamiltonian, $H_{\rm HO} = \hbar\omega a^\dagger a$.
Because the coherent states are eigenstates of the annihilation operator this propagator is simply expressed in this basis \cite{K2}.
\begin{eqnarray}
\int_{A-L_0(0)}^{B-L_0(\tau)}  \D [\delta L]\  e^{iS_{\rm HO}[\delta L]/\hbar}
&=& \langle {B-L_0(\tau)} \vert {e^{-i \omega a^\dagger a t}} \vert {A-L_0(0)} \rangle \pcr
&=& \langle {B-L_0(\tau)} \vert  {{e^{-i \omega t}}(A-L_0(0))} \rangle\pcr
&=& \exp\big\{ -\half \vert B-L_o(\tau)\vert^2 -\half \vert A-L_o(0)\vert^2 \pcr
&&+e^{-i\omega \tau}(B-L_o(\tau))^*( A-L_o(0)) \big\} \nonumber
\end{eqnarray}
Because the expansion about the extreme path left a simple path integral, it is possible to evaluate the integral even though the boundary conditions on the path integral are not the trivial boundary conditions that one expects in a stationary phase approximation.
These non-trivial boundary conditions occur because our extreme path did not satisfy the boundary conditions of the original path integral.

Putting this result into the expression \eq{prop1} for the
propagator the following is found.
\begin{eqnarray}\label{prop2}
\langle {B} \vert U(\tau) \vert {A} \rangle 
&=& \exp \Big\{ {i \over \hbar}S[L_0, M_0^*] - { 1 \over 2}[ L_0^*  \delta L-L_0\delta M^* ]_0^\tau \pcr
&&-\half \vert B\!-\!L_o(\tau)\vert^2 -\half \vert A\!-\!L_o(0)\vert^2 \pcr
&&+e^{-i\omega \tau}(B\!-\!L_o(\tau))^*( A\!-\!L_o(0)) \Big\}
\end{eqnarray}
It is important to note that each of the terms in this expression depends on the choice of extreme paths $M_0$ and $L_0$.  
This is the price to be payed for not having the extreme paths satisfy the boundary conditions of the path integral.  
This indicates that there is a problem, since the propagator should not depend on the freely chosen extreme paths. 
It will be shown that this problem is only apparent.

The first step in disentangling the extreme paths from the propagator is to evaluate the extreme action (the first term in \eq{prop2}).
\begin{eqnarray}\label{CAction}
S[L_0,M_0^*]
&&= \hbar\int_0^\tau dt \ \left\{{i \over 2}\left(\dot L_0 M_0^*-L_0\dot M_0^*\right) - \omega L_0M_0^* - fL_0 - f^*M_0^*\right\}  \pcr
&&=\hbar\int_0^\tau dt \ \left\{{1 \over 2}\left[(\omega L_0 + f^*) M_0^*+L_0(\omega M_0^* + f)\right] - \omega L_0M_0^* - fL_0 - f^*M_0^*\right\}  \pcr
&&=-{\hbar \over 2}\int_0^\tau dt \ \left( fL_0 + f^*M_0^*\right)
\end{eqnarray}
In order to evaluate this further it is useful to write out the solution to the equations of motion.  
The Green's function solution is as follows.
\begin{eqnarray}
L_0(t) &=& e^{-i\omega t} \left[ L_0(t') e^{i\omega t'} - i \int_{t'}^t ds\ e^{i \omega s} f^*(s)  \right]\pcr
M_0^*(t) &=& e^{i\omega t} \left[ M_0^*(t'') e^{-i\omega t''} + i \int_{t''}^t ds\ e^{-i \omega s} f(s) \right] \nonumber
\end{eqnarray}
Both sets of constants $t'$,$t''$ and $L_0(t')$,$ M_0^*(t'')$ are arbitrary.  
Thus, without loss of generality one can choose $t'=0$ and $t''=\tau$, since the free choice of $L_0(t')$ and $ M_0^*(t'')$ will allow any solution to be described.
With this choice the solutions becomes the following.
\begin{eqnarray}\label{CP}
L_0(t) &=& e^{-i\omega t} \left[ L_0(0)  - i \int_0^t ds\ e^{i \omega s} f^*(s)  \right]\pcr
M_0^*(t) &=& e^{i\omega t} \left[ M_0^*(\tau) e^{-i\omega \tau} + i\int_\tau^t ds\ e^{-i \omega s} f(s) \right]
\end{eqnarray}

With this in hand the extreme action can be evaluate more fully.
Putting the extreme path \eq{CP} into the expression for the extreme action \eq{CAction} the extreme action can be written as follows.

\begin{eqnarray}\label{classicalAction2}
S[L_0,M_0^*] 
&=&-{\hbar \over 2}\int_0^\tau dt \ \Bigg\{ f(t)e^{-i\omega t} \left[ L_0(0)  - i \int_0^t ds\ e^{i \omega s} f^*(s)  \right] \pcr
&&+ f^*(t)e^{i\omega t} \left[ M_0^*(\tau) e^{-i\omega \tau} + i \int_\tau^t ds\ e^{-i \omega s} f(s) \right]\Bigg\} \pcr
&=&-{\hbar \over 2}\Bigg\{L_0(0)\underbrace{\int_0^\tau dt \ f(t)e^{-i\omega t} }_g 
- i  \int_0^\tau dt \ \int_0^t ds\ e^{i \omega (s-t)} f(t)f^*(s)  \pcr
&&+  M_0^*(\tau) e^{-i\omega \tau}\underbrace{\int_0^\tau dt \ f^*(t)e^{i\omega t} }_{g^*}
+ i \int_0^\tau dt \ \int_\tau^t ds\ e^{i \omega (t-s)} f(s)f^*(t) \Bigg\} \pcr
&=&-{\hbar \over 2}\Bigg\{L_0(0)g 
- i  \int_0^\tau dt \ \int_0^t ds\ e^{i \omega (s-t)} f(t)f^*(s)  \pcr
&&+  M_0^*(\tau) e^{-i\omega \tau}g^*
- i \int_0^\tau dt \ \int_t^\tau ds\ e^{i \omega (t-s)} f(s)f^*(t) \Bigg\} \pcr
&=&-{\hbar \over 2}\Bigg\{L_0(0)g 
- i  \int_0^\tau dt \ \int_0^t ds\ e^{i \omega (s-t)} f(t)f^*(s)  \pcr
&&+  M_0^*(\tau) e^{-i\omega \tau}g^*
- i \int_0^\tau ds \ \int_s^\tau dt\ e^{i \omega (s-t)} f(t)f^*(s) \Bigg\} \pcr
&=&-{\hbar \over 2}\Bigg\{L_0(0)g +  M_0^*(\tau) e^{-i\omega \tau}g^*
-2 i \underbrace{\int_0^\tau dt \ \int_0^t ds\ e^{i \omega (s-t)} f(t)f^*(s)}_h \Bigg\} \pcr
&=&-{\hbar \over 2}\left[L_0(0)g +  M_0^*(\tau) e^{-i\omega \tau}g^*-2 i h \right] \pcr
\end{eqnarray}

Note that $g$ and $h$ depend only on the function $f(t)$ and the time interval $\tau$.
Also notice that by using \eq{CP} and the definition of $g$, that $L_0(\tau)$ can be written in terms of $L_0(0)$ and $g$.
Similarly $M_0^*(0)$ can be written in terms of $M_0^*(\tau)$ and $g$.
\begin{eqnarray}\label{CPBC}
L_0(\tau) &=& e^{-i\omega \tau} \left(L_0(0)  - i g^*  \right)\pcr
M_0^*(0) &=& M_0^*(\tau) e^{-i\omega \tau} - i g
\end{eqnarray}
This is useful since the propagator \eq{prop2} can now be expressed in terms of these boundary values and the boundary condition that $L$ goes from $A$ to $B$.
By putting \eq{classicalAction2} and \eq{CPBC} into \eq{prop2} and doing a bit of algebra the final result for the propagator is found.

\begin{eqnarray}\label{prop3}
\langle {B}\vert U(\tau) \vert {A} \rangle
&=& \exp \Big\{ -{i \over 2}\left[L_0(0)g +  M_0^*(\tau) e^{-i\omega \tau}g^*-2 i h \right] 
- {1 \over 2}[ L_0^*  \delta L-L_0\delta M^* ]_0^\tau \pcr
&&-{1 \over 2} \vert B\!-\!L_0(\tau)\vert^2 -{1 \over 2} \vert A\!-\!L_0(0)\vert^2 
+e^{-i\omega \tau}(B\!-\!L_0(\tau))^*( A\!-\!L_0(0)) \Big\}\pcr
&=& \exp{1 \over 2} \Big\{ -iL_0(0)g -i  M_0^*(\tau) e^{-i\omega \tau}g^*-2 h  \pcr
&&- L_0^*(\tau) \delta L(\tau)+L_0^*(0) \delta L(0) +L_0(\tau)\delta M^*(\tau)  -L_0(0)\delta M^*(0)   \pcr
&&- \vert B\!-\!L_0(\tau)\vert^2 - \vert A\!-\!L_0(0)\vert^2   
+2e^{-i\omega \tau}(B\!-\!L_0(\tau))^*( A\!-\!L_0(0)) \Big\}\pcr
&=& \exp{1 \over 2} \Big\{ -iL_0(0)g -i  M_0^*(\tau) e^{-i\omega \tau}g^*-2 h  \pcr
&&- L_0^*(\tau) (B\!-\!L_0(\tau)) +  L_0^*(0) (A\!-\!L_0(0))\pcr
&&+L_0(\tau)(B\!-\!M_0(\tau))^*-L_0(0)(A\!-\!M_0(0))^*   \pcr
&&- (B\!-\!L_0(\tau))(B\!-\!L_0(\tau))^* - (A\!-\!L_0(0))(A\!-\!L_0(0))^* \pcr
&&+2e^{-i\omega \tau}(B\!-\!L_0(\tau))^*( A\!-\!L_0(0)) \Big\}\pcr
&=& \exp\left\{  - \half AA^* - \half BB^*+e^{-i\omega \tau}B^*A -ig^*e^{-i\omega\tau}B^*  - i gA - h \right\} \pcr
\end{eqnarray}

Notice that the final result does not depend in any way on the particular extreme paths chosen.
Indeed if this were not the case, and the freely chosen parameters of a computational crutch remained in the final answer, the result would certainly be erroneous.

%%%%%%%%%%%%%%%%%%%%%%%%%%%%%%%%%%%%%%%%%%
\section{Verifying the Result}

It will be easier to verify that this result is correct if the coherent state representations of the creation and annihilation operators are found first.
Since the coherent states are the eigenbasis of the annihilation operator,
$a\vert A \rangle = A \vert A \rangle$ and $\langle A \vert a^\dagger = A^*\langle A \vert $.  
With this, the creation operator can be written in the coherent state representation.
$$
\langle B \vert a^\dagger \vert \psi \rangle
= B^* \langle B \vert \psi \rangle 
$$

The representation of the annihilation operator can be found similarly.
\begin{eqnarray}
\langle B \vert a\vert\psi\rangle
&=& \int dA\ \langle B \vert a \vert  A \rangle\langle  A \vert \psi\rangle \pcr
&=& \int dA\ A \langle B \vert A \rangle\langle  A \vert \psi\rangle \pcr
&=& \int dA\ A\ e^{- \half B^*B - \half A^*A + B^*A} \langle  A \vert \psi \rangle \pcr
&=& \int dA\ \left({\partial \over \partial B^*} + {1\over 2}B \right) e^{- \half B^*B - \half A^*A  + B^*A} \langle  A \vert \psi \rangle \pcr
&=& \int dA\ \left({\partial \over \partial B^*} + {1\over 2}B \right) \langle B \vert A \rangle \langle  A \vert \psi \rangle \pcr
&=& \left({\partial \over \partial B^*} + {1\over 2}B \right)\int dA\  \langle B \vert A \rangle \langle  A \vert \psi \rangle \pcr
&=& \left({\partial \over \partial B^*} + {1\over 2}B \right) \langle B \vert \psi \rangle  \nonumber
\end{eqnarray}
Just as the momentum operator can be written as a derivative in the position basis, 
$p \rightarrow -i\hbar{\partial\over\partial x}$, 
the annihilation operator can be written as a derivative in the coherent state basis. 
$a \longrightarrow {\partial \over \partial B^*} + {1\over 2}B$.

Now the Schr\"odinger equation can be written out in the coherent state basis.
\bea
i\hbar {\partial\over\partial t}\vert \psi \rangle &=& H \vert \psi \rangle \pcr
i\hbar {\partial\over\partial t}\langle B\vert \psi \rangle &=& \langle B\vert H \vert \psi \rangle \pcr
i\hbar {\partial\over\partial t}\langle B\vert U(t)\vert\psi_0 \rangle &=& \langle B\vert H U(t)\vert\psi_0 \rangle  \nonumber
\eea
Since any initial state can be written as a sum of coherent states it can be assume, without loss of generality, that the initial state is a coherent state, 
$\vert \psi_0 \rangle = \vert A \rangle$.  
Thus it is only necessary to show that the following is true
\bea
i\hbar {\partial\over\partial t}\langle B\vert U(t)\vert A \rangle &=& \langle B\vert H U(t)\vert A \rangle \pcr
&=& \langle B\vert \hbar[\omega a^\dagger a + f(t) a + f^*(t) a^\dagger] U(t)\vert A \rangle  \nonumber
\eea
In the above, the hamiltonian that corresponds to the truncated Lagrangian has been used.

Now using the above representation of the creation and annihilation operators we find the following form of the Schr\"odinger equation.
\bea\label{schrodinger}
0 &=& \langle B\vert \left[\omega a^\dagger a + f(t) a + f^*(t) a^\dagger- i {\partial\over\partial t}\right] U(t)\vert A \rangle \pcr
&=& \langle B\vert \left[(\omega a^\dagger + f(t)) a + f^*(t) a^\dagger- i {\partial\over\partial t}\right] U(t)\vert A \rangle \pcr
 &=&  \left[\left(\omega B^*  + f(t)\right) \left({\partial \over \partial B^*} + {1\over 2}B \right) + f^*(t) B^*- i {\partial\over\partial t}\right] \langle B\vert U(t)\vert A \rangle 
\eea
The Schr\"odinger equation is now written out in terms of the matrix elements of the propagator in the coherent state basis.  

Using the propagator \eq{prop3}, the following results are found. 
\be\label{annihilation}
\left({\partial \over \partial B^*} + {1\over 2}B \right)\langle B\vert U(t)\vert A \rangle
= e^{-i\omega\tau}(A-ig^*)\langle B\vert U(t)\vert A \rangle
\ee
\be\label{time}
-i{\partial\over\partial t}\langle B\vert U(t)\vert A \rangle
= e^{-i\omega\tau}\left[if g^* -\omega B^*A + i\omega g^* B^* - f^* e^{i\omega\tau}  B^* - f A\right]\langle B\vert U(t)\vert A \rangle
\ee
Substituting \eq{time} and \eq{annihilation} into \eq{schrodinger}, it is found that the propagator \eq{prop3} satisfies the Schr\"odinger equation.

%%%%%%%%%%%%%%%%%%%%%%%%%%%%%%%%%%%%%%%
\section{Discussion}

In summary, the following has been shown.
\bull The stationary phase approximation can be applied to coherent state path integrals without altering the action.
\bull The paths about which one expands need only to satisfy the equations of motion, they do not need to satisfy the boundary conditions.
\bull The path and it's complex conjugate can be expanded about different extreme paths.
\bull There is not a unique extreme path.

The stationary phase approximation failed for Shibata and Takagi \cite{IJMP} because they did not account for the nonzero boundary term that results from the integration by parts in \eq{boundaryTerm}.  
They altered the boundary conditions, following Klauder \cite{K3}, so that it was possible to find a classical path satisfying these altered boundary conditions.  
But these altered boundary conditions were different from the boundary conditions of the path integral.  
Thus there should have been a nonzero boundary term from the integration by parts in equation 3.18a and the boundary conditions on their remaining path integral (equation 3.18b) should have had non-trivial boundary conditions.
When they found the continuous-time coherent state path integral to give an erroneous propagator, they concluded that this was due to a failure of the continuous-time coherent state path integral.
This was an unfounded conclusion since the failure was due an error with the boundary conditions rather than the formalism.

Klauder's prescription chooses the particular boundary conditions for the extreme paths of $L_0(0)=A$ and $M_0(t)=B$.
It is unclear what is special about this particular extreme path.
This paper has demonstrated that any path satisfying the equations of motion will suffice for the stationary phase approximation.
This casts into doubt the validity of his argument leading to this special case since it picks one of many possible extreme paths as the ``true classical path''.

This also puts into question the usual interpretation of the classical path being the ``most probable" or the one ``that contributes the most to the path integral", for coherent state path integrals.
Since there are many equivalent extreme paths, they cannot all have this interpretation.
The extreme path in coherent state path integrals does not appear to have the same significance as the classical path in position space path integrals.

 A final note: In checking our result we also compared our propagator (for the special case of initial and final states having zero momentum) with an equivalent propagator from Feynman and Hibbs \cite{FeynmanAndHibbs}, equation (8-141).  
We find that there is an error in their result in one of the coefficients:  ${1\over m\sqrt{2\omega}}$ should be ${1\over \sqrt{2\hbar m\omega}}$.

\newpage

\end{document}